# Cassiopée, towards technological development for XAO on ELT: the e-APD infrared detector

Jean-Luc Gach*[a], Piero Bruno[b], Julien Charton[b], Philippe Feautrier[a], Thierry Fusco[c,d], Benoit Neichel[d], Jean-François Sauvage[e,d]

[a]First Light Imaging S.A.S., Europarc Ste Victoire – Bât. 5, Route de Valbrillant – Le Canet, 13590 Meyreuil, France; [b]ALPAO S.A.S., 727 Rue Aristide Berges, 38330 Montbonnot-Saint-Martin, France; [c]DOTA, ONERA, Université Paris-Saclay, 91120, Palaiseau, France; [d]Aix Marseille Univ, CNRS, CNES, LAM, Marseille, France; [e]DOTA, ONERA, F-13661 Salon Cedex Air, France;

## ABSTRACT

The Cassiopée project aims to develop the key technologies that will be used to deploy very high-performance Adaptive Optics for future ELTs. The ultimate challenge is to detect earth-like planets and characterize the composition of their atmosphere. For this, imaging contrasts of the order of $10^9$ are required, implying a leap forward in adaptive optics performance, with high density deformable mirrors (120x120 actuators), low-noise cameras and the control of the loop at few kHz. The project brings together 2 industrial partners: First Light Imaging and ALPAO, and 2 academic partners: ONERA and LAM, who will work together to develop a new camera for wavefront sensing, a new deformable mirror and their implementation in an adaptive optics loop. This paper will present the development of the fast large infrared e-APD camera which will be used in the wavefront sensor of the system. The camera will integrate the latest 512x512 Leonardo e-APD array and will benefit from the heritage of the first-light imaging's C-RED One camera. The most important challenges for the application are the autonomous operation, vibration control, background limitation, compactness, acquisition speed and latency.

## 1. INTRODUCTION

Today, there are four AO instruments specifically dedicated to the study of exoplanets, operating on large 8-10m telescopes. The European instrument SPHERE, developed by ESO with significant contribution from our teams, and three projects in North America (GPI at Gemini, SCExAO at Subaru, and KPIC at Keck). These systems have been operational for nearly a decade, and experience shows that for these four instruments, the two main performance limitations are: (1) fast atmospheric residuals, poorly compensated by AO correction operating at limited speeds, and (2) optical aberrations induced by the telescope (phasing errors, wind shake, internal vibrations, local turbulence generated by temperature differences of various structural elements, etc.), which are poorly estimated by AO systems that were not originally designed to measure and correct these defects. Overcoming these limitations has become essential for the success of the next generation of "planet hunters" and to hope, by 2035, to detect a terrestrial planet (similar to Earth) around a nearby star. This scientific and technical challenge will require the combination of the future European giant telescope, the E-ELT, which is four times larger and sixteen times more powerful than its predecessors, and an AO system (and its components) that is also several tens of times more complex than those currently in operation. This race for performance will directly translate into a need for detectors and deformable mirrors that are 2 to 3 times faster and 16 times larger than the current state of the art in the field.

In this context, the development of the new components proposed by FLI and ALPAO will make it possible to tackle both the problem of speed, and space scaling in anticipation of the ELT. However, the development of these key technological building blocks is not sufficient to achieve the performance required for the detection of telluric exoplanets. Indeed, the best AO systems today achieve contrasts of the order of $10^6$, and the final path to $10^9$ will come from new components, their fine characterization and control, and their integration into a complete AO loop, this is the main goal of the Cassiopée project.

*jeanluc.gach@first-light.fr; phone +33 442612920; https://www.first-light-imaging.com/

The actual path to this is 120x120 deformable mirrors and the corresponding wavefront sensor. The development of a 512x512 fast infrared camera is a logical improvement to achieve such a loop. Detailed system description of the whole AO loop can be found in Neichel et al. [1].

## 2. LEONARDO LM-APDS

The physics of avalanche gain in HgCdTe has been well reported, for instance , and the references therein [2][3][4]. Also the HgCdTe Metal Organic Vapour Phase Epitaxy (MOVPE) technology employed by Leonardo for all its infrared detector products has been well described [6][2]. The design and performance has evolved rapidly in recent years and it is worth summarizing the particular technology now used for Lm-APDs (linear mode avalanche photo diodes) also called e-APDs (electron initiated avalanche photo diodes).

Figure 1 illustrates the history-dependent avalanche gain process in HgCdTe that provides the often-quoted noise-free multiplication mechanism. However, the noise figure relates only to gain fluctuations and does not include the noise due to dark current that might also be enhanced by the high bias voltage. The strategy for good APD design must include bandgap engineering (as illustrated in Figure 1) to suppress these currents.

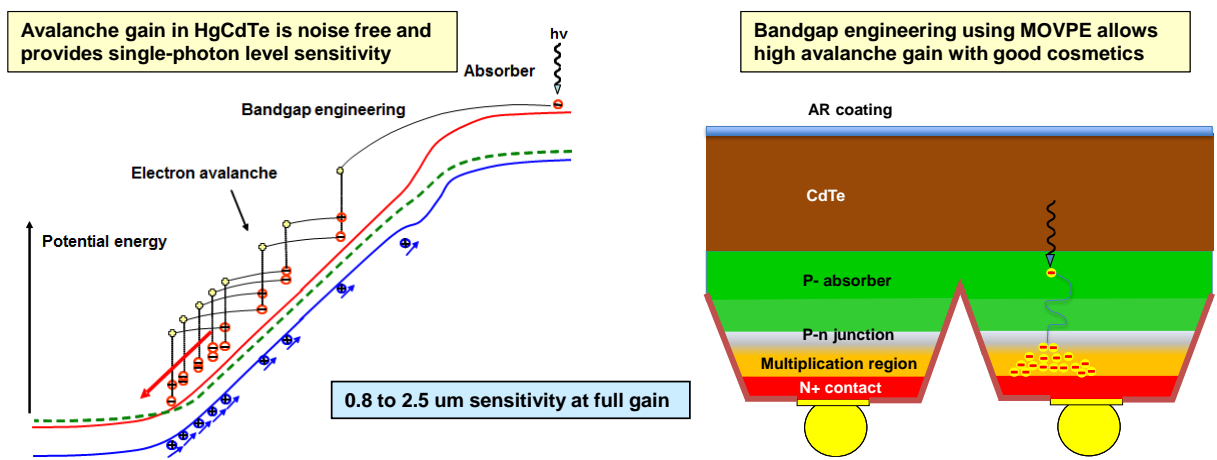

Figure 1:MOVPE design and device schematic (courtesy of Leonardo)

The main advantage of MOVPE for LmAPDs is the ability to incorporate complicated bandgap and doping profiles to switch off junction-related dark current sources, such as: trap-assisted tunnelling and generation-recombination current in the depletion layer. There is also a clear distinction between Liquid Phase Epitaxy (LPE) technologies grown on (111) CdZnTe and MOVPE which is grown close to (100) on GaAs. Homojunctions in LPE are susceptible to 111 threading dislocations and process induced dislocations intercepting the p-n junction. These appear to be absent or much reduced in MOVPE despite growth on a lattice miss-matched substrate. Instead defect revealing etches show a faint pattern of etch pits from weak mosaic crystal boundaries and these can be deactivated by widening the bandgap in critical areas. MOVPE allows for tight geometry control of individual layers so that the volume of narrow bandgap material in the structure is over an order less than planar or via-hole technologies and has a direct effect on thermal current. Bandgap engineering of the LmAPD can ensure that the dark current receives much lower avalanche gain that the photon signal, further suppressing the effect of dark current. In consequence the MOVPE mesa diode is intrinsically less sensitive to dark current mechanisms and pixel defects than other processes. Recent focus has been on bandgap profiles that mitigate tunnel-trap-tunnel currents that ultimately become the key limitation especially at low temperature.

Using an avalanche gain diode permits to maintain a high amplifier bandwidth which is necessary to have high frame rate [6]. Indeed, usually a high bandwidth will end up to a high readout noise not compatible with low light application such as in astronomy. But the avalanche gain will amplify the light induced signal before the readout amplifier (in-pixel source follower in this case), therefore giving a much lower input-referred equivalent readout noise. One can compute the overall input-referred noise using the following equation:

$$\sigma = \sqrt{F \times (QE \times S + S_{Dark}) + \left(\frac{\sigma_{Readout}}{M}\right)^2} \qquad (1)$$

Where F is the excess noise factor occurring usually with gain pixels, in the case of infrared e-APDs it is admitted to be around 1.2. QE is the sensor quantum efficiency, S in the photonic signal in photons, $S_{Dark}$ is the dark signal in electrons, $\sigma_{Readout}$ is the actual amplifier readout noise in electrons and M is the multiplication gain. It is then possible to obtain a very low readout noise while maintaining a high framerate, two antagonist properties. Figure 2 shows the typical input-referred readout noise measured on a Leonardo Saphira device and a C-RED One camera, showing that sub-electron readout noises are achievable with a gain above 20 and a 1700Hz frame rate [9].

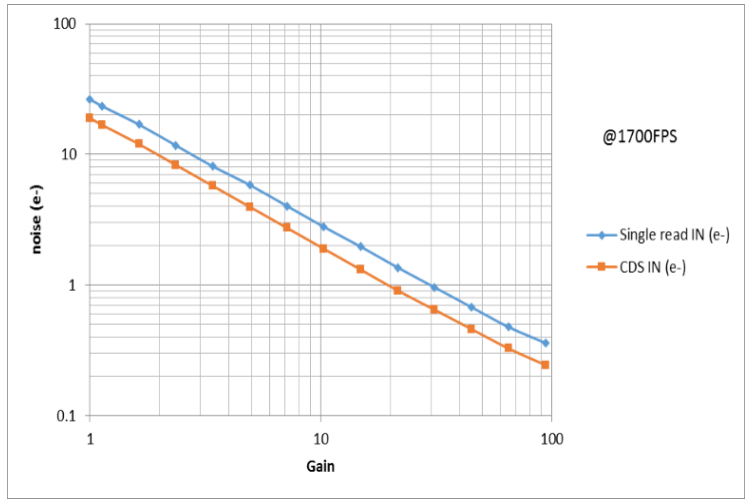

Figure 2: Typical LmAPD input referred noise. Here the performance of a 320x256 saphira device in a C-RED One camera

## 3. THE C-RED ONE HERITAGE

### 3.1 Camera description

C-RED-One is an ultra-low noise infrared camera based on the Saphira LmAPD detector and fabricated by First Light Imaging, specialized in fast imaging camera, after the successful commercialization of the OCAM2 camera [5] dedicated to extreme adaptive optics wavefront sensing. Designed and fabricated by Leonardo UK, the Saphira detector is designed for high speed infrared applications and is the result of a development program alongside the European Southern Observatory on sensors for astronomical instruments [6][7][8]. It delivers world leading photon sensitivity of <1 photon rms with Fowler sampling and high-speed non-destructive readout (>10K frame/s). Saphira is an HgCdTe avalanche photodiode (APD) 320x256 array incorporating a full custom ROIC for applications in the 1 to 2.5µm range. C-RED One camera is an autonomous plug-and-play system with a user-friendly interface, which can be operated in extreme and remote locations. The sensor is placed in a sealed vacuum environment and cooled down to cryogenic temperature using an integrated pulse tube. The vacuum is self-managed by the camera and no human intervention is required. Shown in the Figure 3, the camera has been extensively described in [9].

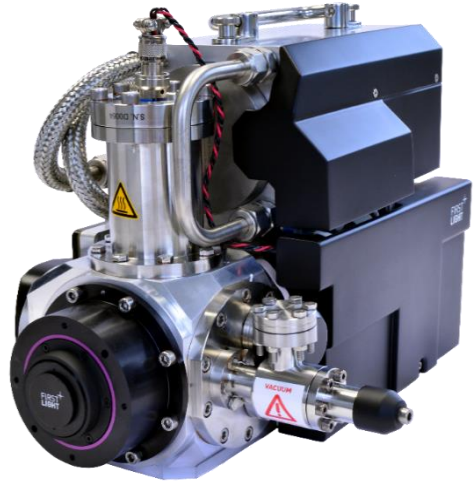

Figure 3: the C-RED One camera. The cooling system (pulse tube) can be seen on the top whereas in the bottom are the vacuum cryostat and the readout electronics.

The Table 1 shows the main C-RED One camera performances.

Table 1: C-RED One 320x256 SWIR APD camera performance

| Test Measurement | Result | Unit |
|---|---|---|
| Maximum speed Full Frame single readout | 3500 | FPS |
| Readout noise at 1720 FPS, CDS readout, gain x50 looking at a blackbody at T=90K | < 1 | e |
| Dark current looking at a blackbody at a temperatire og 90K and gainx10 | < 80 | e/s/pixel |
| Quantization | 16 | bit |
| Quantum Efficiency from 900 nm to 1720 nm at 100 K | ≥ 60 | % |
| Detector operating temperature rage | 80-90 | K |
| Operability due to signal response. Pixels with signal < 0.5*median  at bias of 9V and integration time of 10 ms. | < 0.1 | % |
| Operability due to CDS noise. Pixels with noise > 2*median at bias of 9V and integration time of 10 ms | < 0.1 | % |
| Excess noise factor F | < 1.25 | NA |
| Vibration imparted to the detector, with respect to the front flange of the Camera | < 1 | µm |

## 3.2 Vibration management

The sensor is cooled by a pulse tube. This technology has been developed for intrinsic low vibration and space applications. On top of this, one of the key advantages is a very high reliability of the cooling machine compared to Stirling type which are very common on integrated dewar cooling assembly (IDCA). The pulse tube MTBF is above 90 000h, to be compared to an average of 10 000h for Stirling machines, sometimes less. With Stirling machines, most of the vibrations are caused at the focal plane level by the head piston movment. The vibration level of various Stirling coolers vendors has been reported for example by Kondrajev et al. [10]. The pulse tube cryocooler is quite recent, it works on a thermoacoustic principle and therefore there is only pistons at the compressor level. A relatively small number of scientific publications can be found on them since most of the know-how is being maintained at corporate levels for intellectual property protection. Davis et al. [11] have however published the principles of optimization of such devices. The pistons, actually voice coils, work in linear motion and phase opposition to have a compensating momentum, therefore the residual vibrations are only due to the pistons mass mismatch and the driving phase error, so they are by construction at very low levels. However, this can be even improved by using an active drive mechanism. A MEMS accelerometer is placed on the compressor itself with the sensitivity in the piston displacement direction. The individual driving phase (sinusoidal) of the voice coils is adjusted to null the residual vibration and compensate for the residual mismatches in the compressor construction. This system has been successfully implemented on the C-RED One for the E-ELT phasing and diagnostic station (PDS). The E-ELT is quite sensitive to vibrations. This was identified early on and stringent requirements on any system have been established. Figure 4 shows the force spectrum in 3 directions induced by the camera and the active control improvement. Although the vibration level is already low, using the active control will improve them by a factor of 16.

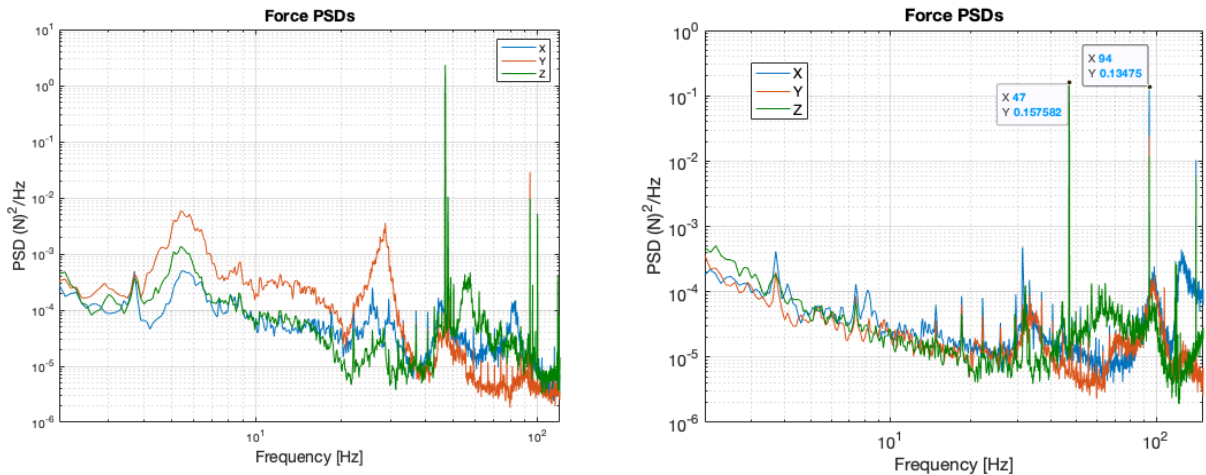

Figure 4: force spectrum of the camera induced vibrations without active control (left) and with active control (right). Note the different (log) scale on the two graphs (measurements courtesy of ESO)

### 3.4 Background management

The quantum efficiency of Leonardo LMAPDs have been measured up to 4 μm wavelength by Finger et al. [12] and it turns out that the sensor is sensitive up to 3.7 μm. The measurement is reproduced here in Figure 5. This is due to the heterostructure of the diode and the gain region which is made of mid wave cutoff MCT. This material is used in the multiplication region to have a large avalanche gain under modest polarization, whereas the absorber region uses shorter wavelength material to minimize the bulk dark current (see Figure 2). This is why also the QE decreases in L band at high gain. Actually QE is the same but the redder photons are not amplified because they penetrate deeply into the MCT material, above the gain region and are not amplified. Therefore, the QE appears to be lower when input referred (divided by the actual gain ratio between red and blue photons). This has been discussed in detail in a previous publication [14].

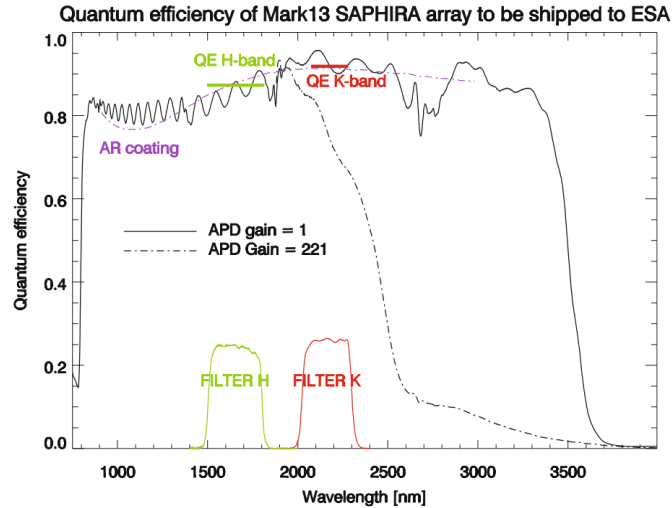

Figure 5: Quantum efficiency of Mark 13 Leonardo LmAPD diodes as reported by Finger et al. in [12].

The sensitivity to long wavelengths is an issue for thermal background management, especially with single photon sensitive devices. It is commonly admitted that working in J and H bands up to 1.8 µm does not require any environmental cooling. This is true for "classical" infrared detectors where the images are limited by the intrinsic readout noise of the detector when it is in the range of 10 to 30 electrons. But with sub-electron readout noise detectors this assumption may not be valid, and the noise performance might be dominated by dark and background induced shot noise (see equation 1). There are several strategies to limit the background induced photonic signal. The first is to put the camera in a cold environment. This is the strategy used by some E-ELT instruments, and a -30°C cooling which implies just dry air to avoid condensation or frost and not a large cryovessel is sufficient to lower the background to negligible levels. The second strategy is to minimize the beam aperture and use cold filters to reject K and L bands photons, this is the standard approach with C-RED One. The final approach is to use cryogenic optics for the final stage and a cold pupil, in that case the seen background is proportional to the source etendue at first order. This is a possible scenario with C-RED One, although more complex and less flexible than the previous one. Most of the users chose the cold filter approach with J+H sensitivity and K+L blocking filters. However, even with blocking filters, the sensor is so sensitive that the small amount of blackbody emission at 293K in H band is detectable and contributes to the overall signal in a significant manner, depending on the camera beam aperture and the longpass cutoff wavelength. This signal, aggregated with dark signal, will also contribute significantly as shot noise to the readout noise at low framerates. Figure 6 shows the simulated effect of the variation of the cutoff wavelength (assuming that the filters are ideal and have an infinite attenuation in the blocking band) and beam aperture on the background collected by the camera. Real measurements were in accordance with these simulations.

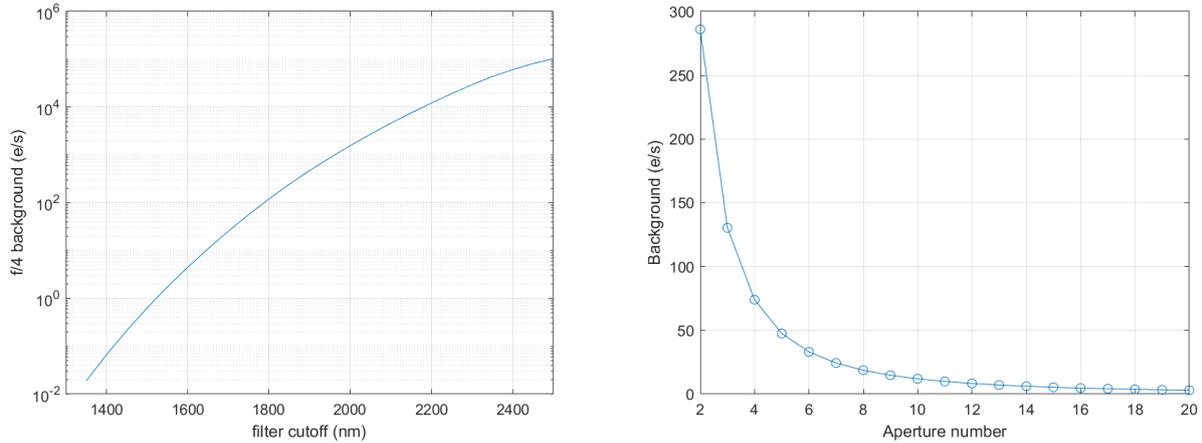

Figure 6: Left, effect of filter cutoff wavelength on the integral background radiation per pixel for a F/4 input beam (293K scene, simulated). Right, effect of the input beam aperture on the integral background radiation for a 1.75mm cutoff wavelength (293 K scene, simulated). In both cases ideal cutoff filters with infinite attenuation are assumed.

Today, with the optimization of the operating voltages to limit the ROIC glow, the background contribution can be far higher that the intrinsic dark current of the sensor. This gives the limits of the blocking filters and hot optics solution, which is very valid in the case of wavefront sensing where the exposure time is usually low (between 100ms and a few 100s of microseconds). This is not the case for longer exposures (several minutes) where a strong spectral bandwidth restriction must be considered or the use of cold optics, either in a cooled enclosure or a hybrid solution with cryogenic final optics and room temperature collimator. In any case a detailed analysis is necessary to ensure the full detector performance, especially in terms of readout noise which might be severely degraded by the background shot noise.

## 4. 512X512 LMAPD DETECTOR

This detector is the result of an ESO funded program and has also been extensively described as it was in the development phase by Finger and his collaborators [12]. Based on the Saphira-ME1001 heritage, a new ROIC has been developed by Leonardo under the name ME1120. The ME1120 ROIC is a 512x512 pixel array designed for low glow with reference pixels. The 512x512 format pixel array on 24μm pitch is compatible with bump-bonded e-APD MCT arrays grown by MOVPE. The ROIC may be configured and controlled using an SPI compatible register-based serial interface. The ME1120 is designed to be a low glow device. Versatile windowing is combined with non-destructive read, read reset read (RRR) and interlaced pixel imaging modes. Nondestructive readout facilitates Fowler sampling to reduce noise and increase sensitivity. Figure Table 2 shows the main detector characteristics and Figure 7 shows a diagram and a picture of the actual detector.

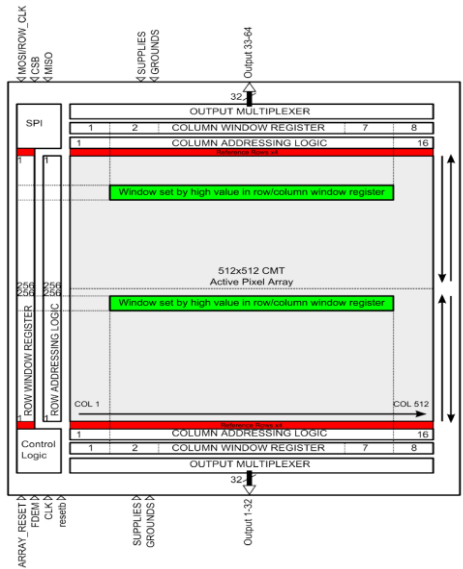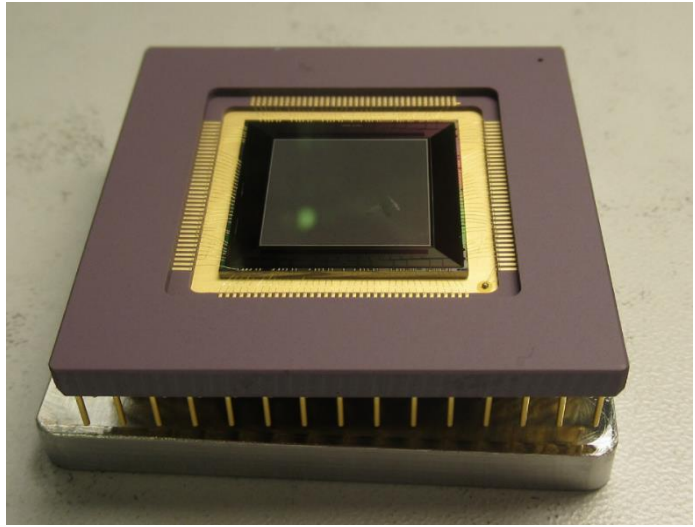

Figure 7 : left, bloc diagram of the ME1120 ROIC. Right, Actual sensor in a PGA package

Table 2: ME1130 FPA full specifications

| Parameter | Value | Units | Comments | Verification |
|---|---|---|---|---|
| Active array size | 512x512 | Pixels | | Design |
| Pixel size | 24 | µm | Square Pixel | Design |
| Fil Factor | >90 | % | Active collection area | Design |
| No of Outputs | 64 | | | Design/Test |
| Read-Out mode | Global Shutter | | | Design/Test |
| Frame Rate | ≥2000 | Hz | At 8.6MHz pixel rate | Design/Test |
| Integration time | Min: 500 Max: no limit | µs | | Design/Test |
| Windowing | | | Multiple window readout – selectable window sizes, flexible in rows but quantized in 32 blocks of 16 columns | Design/Test |
| APD gain | 1-1000 | | Likely to be limited to 100 by the drive electronics | Design/ Characterisation/Test |
| Cut On Wavelength | ≤0.8 | µm | | Design |
| Cut On Wavelength | ≥2.5 | µm | | Design |
| QE x Fill Factor | ≥70 | % | Assumes the appropriate AR coat, between 0.85 & 2.45 µm | Design / Characterisation |
| Defects | 0.2% | | Signal response ± (0.5 × median) | Test |
| Dark Current | <1 | e/p/s | At 90K | Test |

| Read Noise | <1 | e- RMS | In dark conditions. At an APD gain of 20, with CDS | Test |
|---|---|---|---|---|
| Excess Noise Factor | <1.2 | | At 80K, gain less than 30 | Test |
| Full well | >20 | ke- | Defined as the maximum number of charges storable within the linear regime of the detector | Test |
| Non Linearity | <1 | % | up to 80% of full storage capacity | Test / Characterisation |
| Operating Temperature | 80 | K | | Test |
| Operating Voltage | <3.5 | V | except diode array bias | Design |
| Power dissipation | <250 | mW | Full frame readout | Design / Test |

## 5. THE CASSIOPEE CAMERA

The Cassiopée camera will integrate the 512x512 LmAPD sensor from Leonardo, a pulse tube cooling system with a possible active damping and a cold filter plus baffle assembly very similar to the one already used on the C-RED One camera described above. This will permit to use the long heritage acquired already on this camera and limit the development risk. Indeed it turned out from this previous experience that these subsystems are the most critical and sensitive to any change. The baseline of the sensor clocking is 8.6 MHz but it is likely limited by the test equipment at ESO. We expect to raise this at least at 10MHz which is the actual baseline of the C-RED One camera and Leonardo's specification, and maybe even higher. This will raise the frame rate to a minimum of 2.3 kHz in single read full frame, and we expect to reach more than 3kHz. The Table 3 shows expected camera performances.

Table 3: expected Cassiopée LmAPD camera performances

| Parameter | Value |
|---|---|
| Framerate | 3000 FPS single read (goal) 2300 (min), 1500 FPS CDS |
| Readout Noise at 3000 FPS and gain ~50, looking at a black body at a temperature of 80K | < 1 e- (sub e-) |
| Total background (dark current + thermal background) at 3000 FPS and gain ~10 and 80K with 1.7 μm cutoff & F/4 beam, looking at a black body at room temperature | ≤ 30 e-/s/pixel |
| Dark current at gain ~10 and 80K | < 5 e-/s/pixel |
| Quantization | 14 or 16bit |
| Detector Operating Temperature (No LN2) | 80K |
| Quantum Efficiency from 1.1 to 2.4 μm (J,H,K) | >70% |

| Excess noise Factor F | <1.25 |
|---|---|
| Pulse tube vibration imparted to the detector | <1 µm RMS |
| Filtering | Fixed long-wavelength suppression cold filters at front of the camera with cut-off wavelengths around 1.7 µm; |
| Datalink | 2x10GigE or CXP12x2 |

Handling the huge data amount produced by the camera is also a concern. 512×512×2000×16bits gives 8.3 Gbit/s throughput, not taking into account any overhead which is already at the upper limit of a 10GigE link if extra bandwidth is consumed by protocol and data encoding. The baseline then is to use an aggregated 2x10GigE interface or a (not very common) 25GigE link. The first option is preferred aside the industry standard Coaxpress 2.0 which offers up to four 12.5Gbit/s links aggregated (CXP12). We expect to receive the science grade devices by end of 2024 and to build the first H band prototype camera by the end of 2025 which will permit to the consortium to perform the first on-sky tests on the 1.52 m Haute Provence Observatory (OHP) telescope using the PAPYRUS adaptive optics bench developed by Chambouleyron and his collaborators [13]. The commercial availability of the camera will be likely one year later.

## ACKNOWLEDGEMENTS

This project vas funded by the French "4ᵉ programme d'investissements d'avenir régionalisé" (PIA4) under grant number 4076334 named CASSIOPEE: Concepts Avancés et Sous Systèmes InnOvants Pour l'Extremely large tElescope. We would like to thank Ministère de l'Economie, des Finances et de la Souveraineté, BPI France, Région Sud Provence Alpes Côte d'Azur, Région Auvergne Rhône Alpes, for supporting and funding this project.